# A Graph Transformation Strategy for Optimizing SpTRSV

Buse Yılmaz
Dept. of Software Engineering
*İstinye University*
İstanbul, Turkey
buse.yilmaz@istinye.edu.tr

Abdülkadir Furkan Yıldız
Dept. of Software Engineering
*İstinye University*
İstanbul, Turkey
abdulkadir.yildiz@stu.istinye.edu.tr

*Abstract*— Sparse triangular solve (SpTRSV) is an extensively studied computational kernel. An important obstacle in parallel SpTRSV implementations is that in some parts of a sparse matrix the computation is serial. By transforming the dependency graph, it is possible to increase the parallelism of the parts that lack it. In this work, we present an approach to increase the parallelism degree of a sparse matrix, discuss its limitations and possible improvements, and we compare it to a previous manual approach. The results provide several hints on how to craft a collection of strategies to transform a dependency graph.

## I. Introduction

Sparse triangular solve (SpTRSV) is an important building block to several linear algebra systems such as iterative methods [1-2], preconditioners for sparse iterative solvers [2] and least-squares problems [3] and it has been studied extensively. Yet, optimizing SpTRSV remains a challenging problem on parallel architectures due to the following reasons: **1)** Dependencies between computations result in limited parallelism, **2)** Computations are irregular in nature due to the sparsity structure of the matrix: workloads assigned to threads can be fine-grained and vary in granularity, **3)** Due to challenges **1** and **2,** load balancing and synchronization becomes difficult [4].

The challenges stated above arise due to the sparsity pattern of the matrix and it is usually non-uniform throughout the matrix. This results in the sparse matrix having parts exhibiting different degrees of parallelism. And those parts with a low degree of parallelism (having very few computations) pose a bottleneck for parallel SpTRSV algorithms on today's highly parallel architectures: several cores sit idle due to the dependencies. Different approaches have been taken to attack this problem which are based on partitioning the matrix into smaller pieces and solving them with the appropriate method and/or on the appropriate architecture. "Block-diagonal based methods" is an umbrella term which is used for such partitioning methods for SpTRSV on CPUs [5]. In [6], we developed an SpTRSV implementation that distributes the computation between CPU and GPU. Another similar work is presented in [7]. While partitioning approach is natural and has many successful examples in the literature [8-11], in a previous work [12], we proposed exactly the opposite approach. In this approach, the dependency graph of the matrix is transformed to make the sparsity pattern more homogeneous so that the total degree of parallelism of the matrix increases. In other words, we attempt to make parts exhibiting different degree of parallelism similar to each other. Increasing the degree of parallelism in parts where the computation becomes serial is achieved by rewriting the equations that rows represent.

In this paper, we discuss the strategies that can be used in the rewriting approach introduced in [12]. Even if the equation rewriting approach seems straightforward, transforming a directed acyclic graph for the homogeneity of the sparsity pattern is subject to several constraints and there can be countless combinations of the graph nodes depending on how the graph is transformed. Defining and reaching the optimal solution with limitations such as the cost of the transformation, sparsity pattern, how big the graph is, and the underlying parallel architecture, makes this problem quite challenging.

The rest of the paper is as follows: Section II gives background information about SpTRSV and briefly introduces the equation rewriting approach from our previous work [12] for the completeness of the paper. Section III discusses an early and naïve strategy to transform the graph, its limitations, and possible improvements. Section IV presents the experiment results, Section V gives the related work and finally, we conclude in Section VI.

## II. BACKGROUND

### A. Sparse Triangular Solve (SpTRSV)

The sparse triangular solve represents a matrix multiplication operation. It takes a sparse lower triangular matrix $L$ and a right-hand side dense vector b. Then, the linear equation $Lx = b$ is solved for $x$. Fig. 1 presents the operation $Lx = b$ and the dependency graph of the lower triangular matrix $L$ partitioned into levels. Levels are constructed using

This work is supported by the Scientific and Technological Research Council of Turkey (TUBITAK), Grant No. 121E612.

the level-set method [14-18] which is a well-known approach to optimize SpTRSV. Level-set methods are discussed in Section V. For the completeness of the discussion in this section, it should be mentioned that each level is assigned to a thread group and thus the values of the rows in a level can be calculated in parallel. Dependency graph of $L$ is a directed acyclic graph ($DAG_L$), where the nodes represent the rows of L and the edges represent the dependencies between the rows. As seen in the graph, there are no dependencies between the rows within a level. For example, row 7 has 3 incoming edges from rows 0, 3 and 6. These rows correspond to the nonzeros on row 7 in matrix L except for the diagonal (L[7][7]). Hence, the calculation of row 7 depends on values of rows 0, 3 and 6.

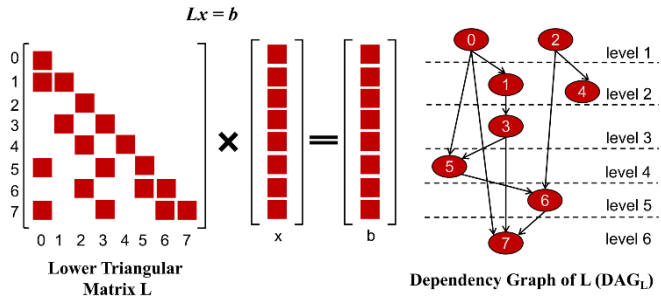

Algorithm 1: Solving SpTRSV serially for $Lx = b$, L is in CSR format
```
for i = 0 to n-1 do
    for j = rowPtrL[i] to rowPtrL[i+1] – 1 do
        sum[colIdx[j]] += valsL[j] * x[colIdx[j]];
    end
    x[i] = (b[i]- sum[colIdx[j]])/valsL[rowPtrL[i+1] – 1];
end
```

Fig. 1 **Left:** $Lx = b$ shown with lower triangular matrix L. L's dependency graph partitioned into levels. **Right:** Algorithm to solve SpTRSV serially for $Lx = b$, where L is in CSR format.

The equation that involves row 7 is as follows:

b[7] = L[7][0] * x[0] + L[7][3] * x[3] + L[7][6] * x[6] + L[7][7] * x[7].

Since Lx = b is solved for x, the unknowns are the values of the x vector. The values of x[0], x[3] and x[6] are calculated beforehand and x[7] is calculated using this equation.

Algorithm 1 on the right hand-side of Fig. 1 presents a serial forward substitution implementation of SpTRSV for solving $Lx = b$. n is the number of rows in L. In the inner for loop, for each row, a partial sum is calculated using the nonzeros (dependencies) in that row except the one on the diagonal. The inner loop can be parallelized for each row, however, the parallelization of the outer for loop is limited by the dependencies: unless all dependencies are met, a row cannot be calculated. This causes a bottleneck on the parallelization of the outer for-loop.

*B. Rewriting The Equations*

As introduced in [12], some equations that the sparse matrix represents are rewritten in order to transform $DAG_L$ to increase the parallelism of the parts that lack it. As a result of the dependencies, these parts usually have very few rows that can be calculated in parallel. Hence, the computational capacity of a parallel architecture is wasted in these parts since the computation is almost serial. Rewriting an equation gives one the liberty to disable or replace the dependencies of the row, causing a shift of the row's node upwards in the dependency graph. In other words, by rewriting the equations, rows can switch levels, changing the granularity of them. This is how levels with very few computations in total – we refer them as *thin levels* – can be "fattened". In addition, the number of levels can be reduced by rewriting the equations of each row in a thin level resulting in the reduction of the synchronization points required. When threads reach to the end of a level, they have to wait at a synchronization barrier: because of the dependencies, the computation of a level has to be finalized before starting the computation of the next level. Hence the computation of levels is parallel within a level but serial across the levels. In this sense, the goal of rewriting the equations is to transform a matrix into a "chubby and short one" where there are several computations in a level that can be done in parallel and there are few levels that are computed one after the other serially. Therefore, rewriting the equations can also be viewed as a way of load-balancing across the levels. As this approach can be coupled with a novel SpTRSV implementation, it can also be used a preprocessing step to transform the matrix, and then existing SpTRSV implementation can be used.

Fig. 2 is taken from [12], showing an example to the equation rewriting method. In this example, equation rewriting is applied to row 3 (x[3]) twice. Row 3 has a dependency on row 1 and row 1 depends on row 0. The dependency graph on the left is the original dependency graph, while the ones in the middle and on the right show the graph after the equation rewriting operations.

Rewriting the equations is basically done by replacing an unknown x (x[1]) in the current equation (x[3]) with its original equation (indicated in blue). First, row 3 is rewritten and it is shifted up from level 3 to level 2 (middle graph), and then the operation is repeated (indicated in green) resulting in row 3 to move from level 2 to level 1 (right hand-side graph). Since level 3 becomes empty, it can be removed. Replacing x[1] with its equation within x[3]'s equation, breaks x[3]'s dependency on x[1] (dotted blue arrow) and now it depends on row 0 (straight blue arrow). The equivalent operation is shown in green in the right hand-side graph.

The equation resulting from the rewriting is:

x[3] = (b[3] - val[3][1] * ((b[1]-val[1][0] * (b[0] / val[0][0])) / val[1][1])) / val[3][3]

It is seen that this equation cannot be represented with a lower triangular matrix. In addition, the number of arithmetic operations increased causing an increase in the computation cost. Hence, this equation needs to be rearranged. This simply can be done by grouping the common multipliers for the unknowns and grouping the constants together. In this equation there are no unknowns since row 3 is shifted to level 1, hence rearrangement should result in x[3] = b[3]'/val[3][3] where b[3]' is all the constants calculated and reduced down to one constant except val[3][3]. This rearrangement is not implemented in [12].

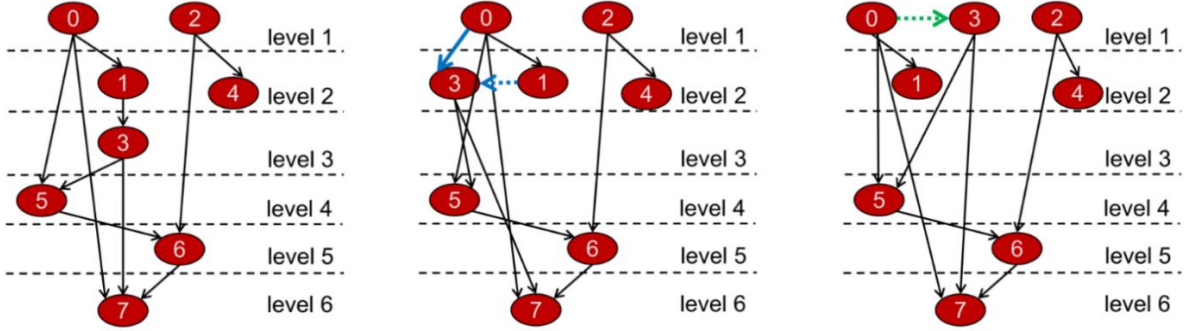

Fig. 2 **Left:** Original dependency graph, **Middle:** Row 3 is rewritten to level 2. Dependency of row 3 to row 1 is broken, new dependency between row 3 and row 0 is established, level 3 is left empty. **Right:** Row 3 is rewritten to level 1. Dependency of row 3 to row 0 is broken.

### III. A NAIVE STRATEGY TO TRANSFORM THE GRAPH

Transforming the dependency graph to increase the parallelism is a daunting task since the possibilities of transformations grow exponentially when large number of rows are rewritten.

In this section, a simple graph transformation strategy is presented along with its limitations and potential improvements. In [12], the graph transformation was done with manually selected rows. The strategy we present here automizes the transformation process.

For rewriting the equations, there are two major aspects of graph transformation: the rows to be rewritten and the levels to which the selected rows will be rewritten. For simplicity and to be able to decrease the number of synchronization barriers, instead of individual rows, *thin levels* can be chosen to be rewritten. Then, a constraint is needed to decide that a level is *thin*.

Another constraint that needs to be set is the stopping criteria. A row can be rewritten until the very first level but each row needs to be rewritten to a level where the transformation process increases the parallelism with a reasonable transformation cost. When the total cost of the level that a row is rewritten to reaches to some threshold the rewriting process for this row should end. The number of dependencies of a row and the number of levels the row is shifted upwards play an important role in the transformation cost. The cost of the graph transformation process needs to be taken into consideration in addition to the performance improvement it provides.

Finally, the underlying architectural features can help with setting additional constraints. At the right hand-side of the equations, each dependency causes a memory access to the x vector. Then these accesses need to be as close to each other as possible to increase the spatial locality. A threshold depending on the cache features of the underlying architecture can be set so that if the memory accesses fall far from each other or the access pattern becomes random altogether, the algorithm can decide not to rewrite a particular row or to end the rewriting process for that row.

In the manual approach [12], a simple and manual method is used, discarding the criteria discussed so far in this section. By examining the dependency graph, the levels with the fewest rows are selected by hand such that when grouped together by the rewriting process, they will have a number of rows similar to the levels with more rows in them. Whole levels are rewritten rather than selecting individual rows from levels. Simply, every 9 levels is rewritten to the $10^{th}$ level. Hence, the levels close to each other are prioritized to form groups to cut on the rewriting cost.

Before presenting the graph transformation algorithm, let us define the cost of a row and cost of a level. The cost of a row is the arithmetic operations (FLOPS) that needs to be done to calculate it. For example, both x[1] and x[3] in Fig. 2 has a cost of 3. After rewriting x[3], it has a cost of 1 if its equation is rearranged. Hence, the cost of a row is $2 * nnz - 1$ and cost of a level is $2 * \sum nnz - n$ where nnz is the number of nonzeros in the level and n is the number of rows in a level.

To choose the rows to be rewritten and the levels to which the selected rows will be rewritten, the algorithm uses the average level cost (*avgLevelCost*) where it is defined as $(\sum (2 * \sum nnz - n))/(num.\ of\ levels)$. The algorithm applies the rewriting process among the *thin levels*. *Thin levels* are the levels with a total cost smaller than *avgLevelCost*. Starting from the second level of the *thin levels* (level 1 is the target level), these levels (source levels) are rewritten to upper levels until the cost of the level to which we are rewriting to (the target level) reaches *avgLevelCost*. Hence, source levels are deleted to reduce the number of levels and target levels' costs are increased to *avgLevelCost*. *avgLevelCost* is kept fixed throughout the process rather than being updated whenever a row is rewritten.

This approach requires that we know a row's possible cost at upper levels. We start from the level that a row originally resides in and calculate its possible costs in the upper levels until we reach a target level: the process starts by selecting level 1 as a target level and it calculates the costs of rows starting from level 2 and stops when the cost of level 1 reaches to *avgLevelCost*. Upon arriving at some level n, the process restarts by selecting level n as the new target level. The goal is to keep source and target levels as close to each other as possible to reduce the rewriting cost. We name the number of levels between the source and the target levels as *rewriting distance*.

An example is shown below. The costs are indicated with red. 4 levels of a DAG are shown where the column "Rows" gives the initial setting: the row number and its cost in red. CostMap shows rows and their costs in levels other than their original levels. Target levels are indicated in grey filling (levels n and n+2). Assuming that there are m levels and *avgLevelCost* is 22, costs of rows of levels n+1 and row 4 of level n+2 in level n are calculated and they are put into the costMap of level n. The rewriting process will stop here for level n since the total cost reaches *avgLevelCost*. Hence, level n will have rows 0,1,2,3 and 4 with a total cost of 21. Level n+1 is left empty and the next target level becomes level n+2. Costs of the rows of level n+3 for level n+2 are calculated and they are placed into the costMap as well. The rows that are rewritten are indicated with a diagonal strike-through.

| Level | Rows | | Cost Map (row #, cost) | | |
|---|---|---|---|---|---|
| n | 0, 1 | 1, 1 | 2, 6 | 3, 6 | 4, 7 |
| n+1 | 2, 3 | 3, 3 | | | |
| n+2 | 4, 3 | 5, 5 | 6, 5 | 7, 7 | |
| n+3 | 6, 3 | 7, 3 | | | |

After the rewriting process, the number of the levels is decreased by 2 and the levels look like:

| Level | Rows | | | | | Level Cost |
|---|---|---|---|---|---|---|
| n | 0, 0 | 1, 0 | 2, 7 | 3, 7 | 4, 7 | 21 |
| n+1 n+2 | 5, 5 | 6, 5 | 7, 7 | | | 17 |

Currently, the implementation is serial, but it can be parallelized easily. Multiple threads can be assigned to calculate the costs of the rows to be rewritten in upper levels. One thread can insert all these rows into the target level so that no locks are needed. Since empty levels are not erased until the process is finalized, the rewriting process can also be started concurrently at the beginning and at the end of the graph.

*A. Limitations*

In the naïve algorithm we present, the stopping criteria for rewriting is set as *avgLevelCost*. At first, this seems to be the correct choice: *thin levels* are the levels with a cost smaller than *avgLevelCost*, and the rewriting process will increase the cost of these levels up to *avgLevelCost* by rewriting the rows in them. But *avgLevelCost* depends on the sparsity pattern and it can cause a row to be rewritten to a very far away level if the cost of the level being rewritten is very small compared to *avgLevelCost* and the number of indegrees (dependencies) do not increase much with rewriting. Rewriting to far away levels will incur a significant rewriting cost. In addition, with many indegrees, precision problems can emerge since rewriting can cause long chains of calculations and the magnitude of the constants in the calculations will increase dramatically. Even with a few indegree values, if the number of indegrees keep increasing, a row can end up having many calculations. The current algorithm does not take the change in the number of indegrees (dependencies) into account while rewriting a row. A related issue will be the divergence between the memory accesses to the dependencies. Depending on the sparsity pattern, the possibility of having random memory accesses can increase with increasing number of dependencies.

Another limitation with the algorithm presented is that entire levels are rewritten. While this is common sense and easy for levels with very few rows, rewriting individual rows will unfold new possibilities in rewriting approaches: additional constraints can be set for the rewriting process. A few examples are: **1)** rewrite if row's indegree < $\alpha$, **2)** rewrite if row is on critical path, **3)** rewrite if distance between indegrees < $\beta$.

Before concluding this section, it is worth mentioning that although the details are left out in this work, the rearrangement of the rewritten equations mentioned in section II, B. are implemented.

IV. EXPERIMENT RESULTS

This section compares the current approach with the manual approach introduced in [12], and when no rewriting is applied. The matrices chosen for this experiment are lung2 and torso2 from SuiteSparse Matrix Collection [13]. lung2 has 109, 460 rows(cols) and 492, 564 nonzeros. When level sets are constructed, lung2 has 479 levels (478 synchronization barriers) and 94% (453 out of 479) of these levels have only 2 rows, hence the computation is serial. torso2 has 115,967(cols) and 1,033,473 nonzeros. torso2 has a triangular shape in terms of number of rows in a level.

TABLE I. COMPARISON OF STRATEGIES

| lung2 | no rewriting | avgLevelCost | manual approach [12] |
|---|---|---|---|
| num. of levels | 479 | 23 **(95% -)** | 67 **(86% -)** |
| avg. level cost | 914.054 | 18938.06 **(20.71x)** | 6520.42 **(7.13x)** |
| total level cost | 437,834 | 435,588 **(~1% -)** | 436,868 **(~1% -)** |
| Size of code (MB) | 9.7 | 8.6 **(11% -)** | 9.5 **(2% -)** |
| num. of rows rewritten | - | 1304 **(1%)** | 898 **(0.8%)** |
| torso2 | | | |
| num. of levels | 513 | 341 **(34% -)** | 284 **(45% -)** |
| avg. level cost | 2014.559 | 3086.443 **(1.53x)** | 5070.183 **(2.51x)** |
| total level cost | 1,035,484 | 1,052,477 **(0.2% +)** | 1,439,932 **(40% +)** |
| Size of code (MB) | 21 | 21 **(0% -)** | - |
| num. of rows rewritten | - | 14655 **(13%)** | 18147 **(16%)** |

This paper introduces a rewriting strategy, but a rewriting strategy only transforms the dependency graph. An SpTRSV implementation is needed to use the transformed graph. This can be any SpTRSV implementation from the literature, but we used our own parallel SpTRSV implementation from [12] as a testbed for the experiments. It generates specialized code for the input sparse matrix. The

```
void calculate0(double* x) {     void calculate1(double* x) {
    x[0] = 1 / 1;                    x[2] = (0.0579356-((-2.93613e-07) * x[0]+(0.0579358) * x[1]))/9.6701e-08;
    x[1] = 85.7849 / 85.7849;        x[3] = (-163.137-(-248.922) * x[1])/85.7849;
}                                }
```

```
void calculate0(double* x) {     void calculate1(double* x) {
    x[0] = 1;                        x[229] = -5.52559e+52 - -5.52559e+52 * x[1];
    x[1] = 1;                        x[230] = 1.30523e+62 - -2.9483e+55 * x[0] - 1.30523e+62 * x[1];
    x[2] = 1;                        x[231] = -1.60336e+53 - -1.60336e+53 * x[1];
    x[3] = 1;                        x[232] = 3.96402e+62 - -8.95193e+55 * x[0] - 3.96402e+62 * x[1];
    x[4] = 1;                        x[233] = -4.65247e+53 - -4.65247e+53 * x[1];
    x[5] = 1;                        x[234] = 1.20387e+63 - -2.71807e+56 * x[0] - 1.20387e+63 * x[1];
    x[6] = 1;                        x[235] = -1.35001e+54 - -1.35001e+54 * x[1];
    x[7] = 1;                        x[236] = 3.65612e+63 - -8.25286e+56 * x[0] - 3.65612e+63 * x[1];
    x[8] = 1;                        x[237] = -3.91731e+54 - -3.91731e+54 * x[1];
```

```
void calculate0(double* x) { void calculate1(double* x) {
    x[0] = 1;                      x[24] = (0.05793555778-((-2.936128058e-07) * x[22]+(0.0579357547) * x[23]))/9.670098234e-08;
    x[1] = 1;                      x[25] = (-163.1371892-(-248.9221327) * x[23])/85.78494353;
    x[2] = 1.000000;               x[26] = 3557579.777222 - -9.219093 * x[22] - 3557587.996315 * x[23];
    x[3] = 1.000000;               x[27] = -7.419860 - -8.419860 * x[23];
    x[4] = 1.000000;               x[28] = 15846392.752971 - -27.991896 * x[22] - 15846419.744868 * x[23];
    x[5] = 1.000000;               x[29] = -23.431905 - -24.431905 * x[23];
    x[6] = 1.000000;               x[30] = 62752046.033024 - -84.991683 * x[22] - 62752130.024707 * x[23];
    x[7] = 1.000000;               x[31] = -69.894048 - -70.894048 * x[23];
    x[8] = 1.000000;               x[32] = 233008018.454555 - -258.059907 * x[22] - 233008275.514462 * x[23];
```

Fig. 3  Code generated with different rewriting strategies for level 0 and level 1. Only first ten lines of code is shown. **Top:** Original equations. Rewriting strategy is not applied (no rewriting). **Middle:** Average level cost is used as the stopping criteria (*avgLevelCost*) **Bottom:** Manual implementation from [12] with a rewriting distance of 10 (manual approach [12]). Every 9 level is rewritten to the $10^{th}$. Equations are rearranged for both *avgLevelCost* and manual approach [12].

implementation is still a prototype, it is in an early stage of development. Hence, the details are left out and no runtime results are presented here as they are irrelevant of the strategy discussed.

Table I presents a comparison of the strategy presented in this paper (based on *avgLevelCost)* with the manual approach from our previous work [12] and the original dependency graph when rewriting is not applied. The table presents several metrics. For lung2, the number of levels drop by 95% by *avgLevelCost* and 86% by the manual approach. The manual approach is hand-written hence it is not an automatic process. It rewrites every 9 levels to the $10^{th}$. It rewrites whole levels as *avgLevelCost* but the decision of the levels to be rewritten is left to the programmer: the number of levels that are rewritten are 44 fewer, resulting in 584 fewer rows to be rewritten. Also, the target levels differ. For example, in *avgLevelCost*, the first 114 levels are rewritten to level 1 whereas in the manual approach only levels [2,10] are rewritten to level 1. In addition, *avgLevelCost* has the freedom to write the rows of a level to different target levels. The average cost of a level differs greatly between the two approaches, and they are 20.71x (*avgLevelCost*) and 7.13x (manual approach) of the no rewriting case. But the total cost of all levels combined are very close to each other for the two approaches as well as the no rewriting case. The reason for this is that the number of dependencies stayed the same most of the time when the rewriting approach is applied to the dependency graph. This is due to the sparsity pattern of lung2: the number of indegrees(dependencies) does not exceed 2 for the rows when they are rewritten. Compared to the no rewriting case, the total level cost is reduced by a small fraction for both approaches. The reduction comes from two sources: the rows that are rewritten to level 1 and for rewritten rows with an indegree greater than 1, the division operation is removed (the calculation is made while rewriting), reducing its cost by 1. The dependencies become zero for the rows that are rewritten to level 1 since there is no computation left to be done. Due to the removal of the division operation and reduction in the number of levels, the size of the generated code is smaller but still very close to the no rewriting case.

Compared to lung2, torso2 has many rows in a level and the its variation is much less across levels. torso2 doesn't have a long chain of very thin levels consisting of 2 rows as lung2. This reduces the number of levels that has a cost lower than *avgLevelCost*. Therefore, the reduction in number of levels is less compared to lung2 for both approaches. Since levels have similar number of rows, in the manual approach, it is difficult to decide on the levels to be rewritten by hand. Therefore, we picked all levels with a cost smaller than *avgLevelCost* and rewrote every 9 level of these to the $10^{th}$ level. The increase in average level cost is much less compared to lung2: 1.53x for *avgLevelCost* and 2.51x for the manual approach. The total level cost slightly increased for both approaches. This is due to the increase in number of

```
void calculate0(double* x) {
    x[0] = 1 / 1;
    x[1] = 85.7849 / 85.7849;
    x[2] = (0.0579356 - (-0.000000*1.000000 + 0.057936*1.000000)) / 9.6701e-08;
    x[3] = (-163.137 - (-248.922133*1.000000)) / 85.7849;
    x[4] = (0.0579356 - (-0.000000*((0.057936 - (-0.000000*(1.000000) + 0.057936*(1.000000))) / 0.000000) + 0.057936*((-163.137189 - (-248.922133*(1.000000))) / 85.784944))) / 9.6701e-08;
    x[5] = (-163.137 - (-248.922133*((-163.137189 - (-248.922133*(1.000000))) / 85.784944))) / 85.7849;
```

Fig. 4  The code generated for the manual approach from [12] when the equations are not arranged (unarranged version of the middle code snippet from Fig. 3).

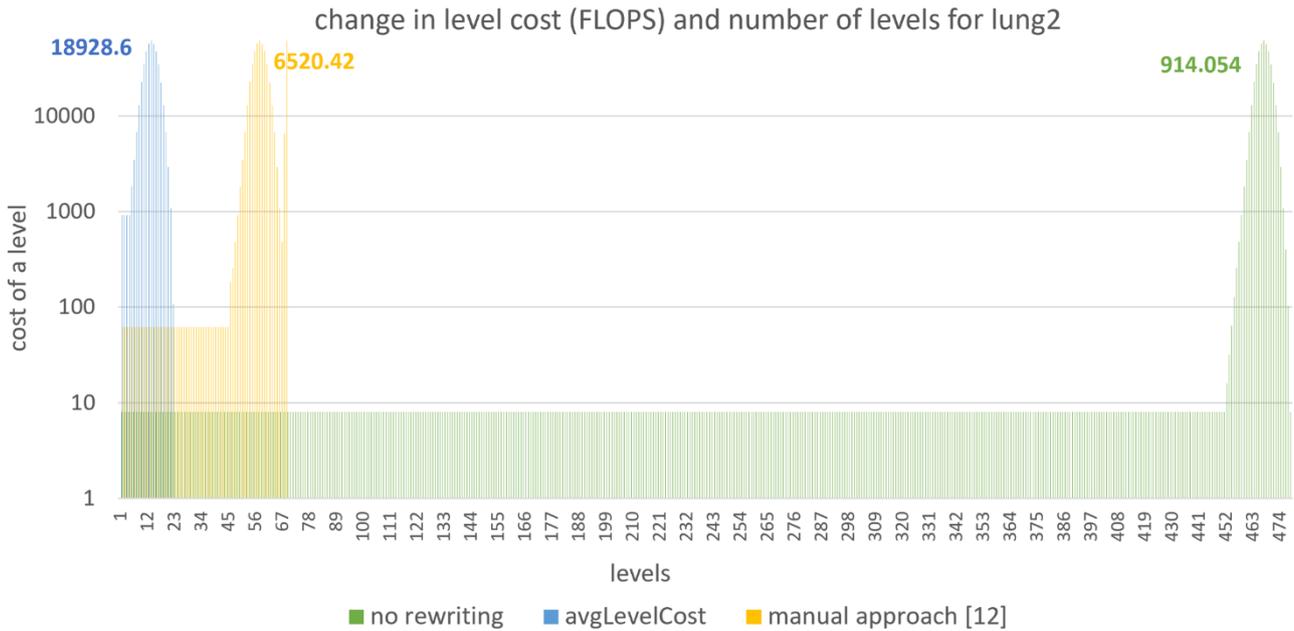

Fig. 5 The number of levels, and the cost of each level in logarithmic scale are provided for the two rewriting strategies presented as well as the case when no rewriting is applied. The average level cost for each three are also provided on top of the graphs.

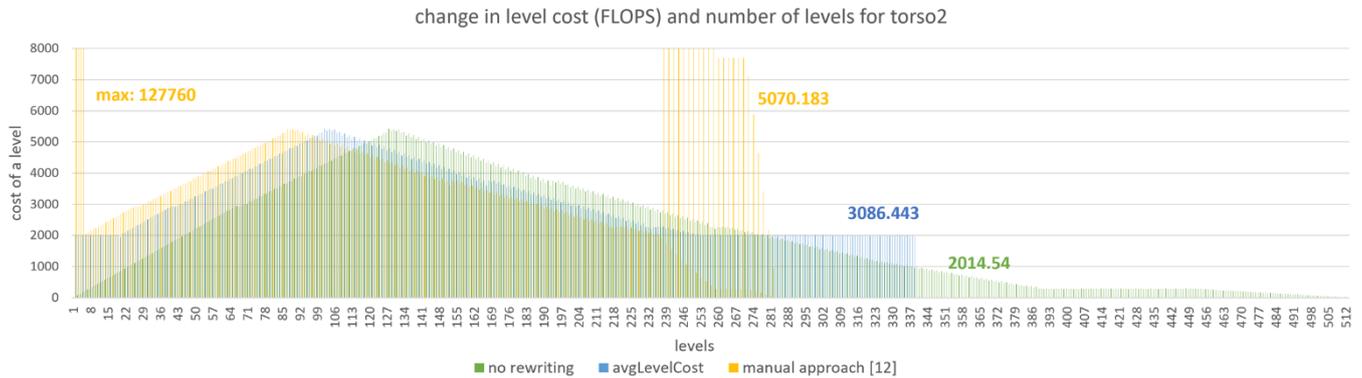

Fig. 6 The number of levels, and the cost of each level for torso2 are provided for the two rewriting strategies presented as well as the case when no rewriting is applied. The average level cost for each three are also provided on top of the graphs.

indegrees(cost of a row) for some of the rewritten rows and stayed the same for the majority. This observation is an appropriate example on the importance of the sparsity pattern and how it affects the rewriting process. Still, the increase in the total level cost is acceptable for avgLevelCost: 2% increase but the manual approach has a 40% increase due to being almost completely blind to the sparsity pattern. The size of the generated code is much more compared to lung2 but it stayed the same as the original for *avgLevelCost*. Since it took a long time, we could not finalize the code generation process and obtain the code size for the manual approach. This can be due to the code generator running serially and without any optimizations: torso2 has twice as many nonzeros as lung2 and the connectivity of the graph (number of indegrees) is much higher making both the rewriting and code generation processes take long time. The manual approach for torso2 is rewriting over 18000 rows which is 16% of the matrix.

Fig. 3 shows the code snippets generated by our SpTRSV implementation for lung2. Only the first 10 lines of the code generated for the first two levels are shown side by side. At least one function is generated for each thread, long levels are divided into multiple functions. The code snippets presented belong to a parallel implementation, but the code generator can choose to generate code to calculate a level with only 1 thread if there are not enough calculations for multiple threads. This is decided by the load balancing mechanism implemented in the code generator.

In Fig. 3, the top snippet shows the original equations, therefore, the rewriting strategy is not applied. The middle snippet shows the top 2 levels when the naïve algorithm presented in this paper is used. The bottom snippet shows the top 2 levels when the manual implementation from [12] is used. In the middle snippet, 227 rows in total are rewritten to level 1 whereas in the bottom snippet, only 22 rows are rewritten to level 1. In the middle snippet, it is observed that the unknowns on the right hand-side of the equations in the rewritten rows became very large in magnitude which affects the precision and accumulates as error for some x values. Hence, this shows us that rewriting process can affect numerical stability if it is overdone: the rewriting distance should be kept small enough so that it does not cause wrong calculations.

The equations shown in the code snippets from Fig. 3 are rearranged into Lx = b format. For a comparison, the unarranged version of the bottom code snippet for the manual approach [12] from Fig. 3 is shown in Fig. 4. There are no dependencies for the rows since level 1 is shown, but the equations are very crowded and they are not in Lx = b format. This code from [12] wastes cpu cycles by doing the same computations over and over for different unknowns. In the current implementation (Table I and Fig. 3), this is fixed by rearranging the equations into Lx = b format while rewriting the rows.

Fig. 5 and Fig. 6 compare the two approaches as well as the no rewriting case graphically in terms of the number of levels and the cost of levels. In Fig. 5, the flat parts represent the levels that are thin, the rows in these levels are rewritten. The length of the flat parts differ among the approaches since the number of rows rewritten are different from each other. However, the bumps are the same since those are fat levels, and the rows in those levels are not rewritten in neither of the approaches. Hence, these parts are already suitable for parallel computation. The y-axis represent the cost of levels and it is in logarithmic scale. The average level cost for each three graphs is also shown on the top. In Fig. 6, we chose not to put the y-axis in logarithmic scale as it is not very representative of the differences between the approaches. The graph shows Instead, we cut the graph at 8000 and indicated the maximum FLOPS in a level for the manual approach. As seen, the manual approach increased the number of FLOPS in an illogical way for some levels causing a great increase in both the average and total level cost.

The results show that the rewriting approach is an effective way to reduce the number of synchronization points (the critical path) and to increase the number of rows that can be computed in parallel (rows of a level). The rewriting process can increase the parallelism of a sparse matrix part where the computation is sequential, and still keep a total cost similar to the original. A dependency graph can be transformed in several ways and some constraints are needed to guide the transformation process. The transformation cost should be kept as low as possible by carefully deciding on the rewriting distance and maintaining the accuracy of the computations.

## V. RELATED WORK

The most well-known body of approaches to optimization SpTRSV are level-set methods [14-18] and synchronization-free methods [19-23]. In level-set methods, rows that have no dependencies to each other are grouped into levels. Hence, rows within a level can be calculated in parallel using a group of threads. Since rows in different levels can have dependencies to each other, at the end of each level, a synchronization barrier is needed and the levels are computed one after the other, in a serial manner. As mentioned in Section II, this work uses level-set method too.

Even though level-set methods are more appropriate for CPUs, both CPU and GPU implementations exist. In [16], level-set based method with matrix reordering is implemented for SpTRSV on CPUs. Due to the fine-grained nature of the computations in SpTRSV, dynamic scheduling on GPUs tend to performing worse compared to CPUs [16]. Later, this approach is adopted for GPUs in [17] and in [18].

Although level set methods partition the dependency graph horizontally in accordance to the natural ordering of the rows, the levels incur overhead due to synchronization barriers and they have to be computed serially. Furthermore, all threads need to reach the barrier more or less at the same time, otherwise they will have to wait idle for each other. This scenario might indicate an ineffective load balancing approach but more importantly, rows whose computation is finished cannot "unlock" the computations for the rows depending on them right away due to the barriers. The sparsity pattern of the matrix can complicate load balancing further. In a previous work [4], we have shown how the number of threads active in a level can fluctuate throughout the levels. Unless partitioning of the rows are made with some another approach such as the equation rewriting approach, some threads tend to stay idle within thin levels.

Synchronization-free methods emerged as an alternative to level-set methods to eliminate these drawbacks. Instead of horizontally, synchronization-free methods group rows according to dependencies, hence the partitioning is vertical. First examples of this approach were on CPUs in [19]. Later, several implementations on GPUs [20-23] have been proposed. In [21], authors introduce a counter-based scheduling mechanism and use a parallel topological sorting algorithm to set the levels and each element only waits for its own dependencies. In [22], authors propose a simple preprocessing phase, where self-scheduling mechanism is set up based on the in-degree of dependency graph nodes. With synchronization-free methods, fine-grained tasks can be created, and synchronization barriers are eliminated but this method requires hundreds or thousands of threads to be assigned to tasks and do busy-waiting on their predecessors. Hence, synchronization-free methods are seen more suitable for GPUs.

Although a few, hybrid implementations of level set and synchronization-free methods exist in the literature [4], [24]. In [24], such a hybrid approach is developed for CPUs. In our previous study [4], we adopted a similar approach to [24]. Both works aim to reduce synchronization points for SpTRSV on CPUs, eliminating unnecessary dependencies.

In addition to level-set methods and synchronization-free methods, other approaches are also available for SpTRSV. One of the most common optimizations is to reorder the sparse matrix and the dense vectors to increasing cache locality. Graph-coloring is another approach that is used for the optimization of SpTRSV. It is utilized as a load balancing approach. Early examples are on CPU by [25-26]. Later, this approach is adopted for GPUs also [27-28]. Graph-coloring is an NP-Complete problem, and it usually requires matrix reordering which may require additional preprocessing.

Block-diagonal based methods form another approach to improve SpTRSV. It is a method practiced on CPUs more than compared to GPUs. Matrix is reordered to help forming blocks around the diagonal and off-diagonal to improve the locality [8-9,29-32]. In [8], authors use dense BLAS operations to compute dense off-diagonal blocks. Dense matrix computations have higher parallelism than their sparse counterparts, therefore, creating dense blocks in sparse matrices is an effective method to improve performance. In [30], authors introduce a code generation framework where it generates SpTRSV code for various sparse matrix operations. One of the code transformations is to convert a sparse code to a set of non-uniform dense sub-kernels. Efficiency of these techniques are highly dependent on the sparsity structure of the matrix.

## CONCLUSION

In this paper, we proposed an automated early algorithm to transform the graph and discussed the constraints that the transformation process is subject to. In addition, we discussed the limitations of the algorithm and compared it to a manual approach from our previous work [12]. Our next goal is to incorporate the constraints discussed in the paper into the algorithm and work on the limitations. Due to the large possibilities of transformations, we are planning to develop the transformation process into a more sophisticated approach such as a multi-objective algorithm or reinforcement learning algorithm. In the long term, we are aiming to create a collection of graph transformation strategies which can be applied in a stand alone manner as well as in combination.